\documentstyle[preprint,aps,eqsecnum,epsfig]{revtex}
\tighten
\begin{document}
\preprint{LPTHE-98-  UCMFTEOR-98- , DEMIRM-98-}
\draft \title{\bf GENERATION OF GRAVITATIONAL WAVES BY GENERIC SOURCES IN 
DE SITTER SPACE-TIME}
\author{{\bf H. J. de Vega$^{(a)}$, 
J. Ramirez $^{(b)}$, N. Sanchez$^{(c)}$ }}
\address { (a) LPTHE\footnote{Laboratoire Associ\'{e} au CNRS UMR 7589.} 
Universit\'e Pierre et Marie Curie (Paris VI) et Denis Diderot (Paris VII),
Tour 16, 1er. \'etage, 4, Place Jussieu 75252 Paris, Cedex 05, France \\
(b) Departamento de Fisica Teorica I. Universidad Complutense de Madrid.
28040 Madrid. Spain. email: mittel@fis.ucm.es \\ 
(c) DEMIRM. Observatoire de Paris. }
\date{\today} 
\maketitle

\begin{abstract}
We study the generation of gravitational radiation by sources
moving in the de Sitter background. Exploiting the maximal symmetry and
the conformal flatness of de Sitter space-time we prove that the 
derivation
of this gravitational radiation can be done along the same lines as in 
Minkowski space-time. A gauge is chosen in which all the physical and
unphysical modes of the graviton are those of a minimally coupled
massless scalar field in de Sitter space-time
 and a massless field in Minkowski space-time.
The graviton retarded Green's function and the Schwinger commutator function
are computed in this gauge using Quantum Field Theory techniques.
We obtain closed formulae 
for the spectral decomposition in frequencies of the linearized gravitational
field produced by the source, in terms of a suitable spectral decomposition
of the source energy-momentum tensor $T^{(1)\nu}_{\mu}$. This spectral
decomposition is dictated by the free (sourceless) gravitational wave modes
in the de Sitter background. 
   
\end{abstract}


\section{Introduction}
De Sitter space-time is interesting for several reasons. 
To begin with, the de Sitter 
space-time is the natural framework for inflationary models of the universe, 
in which the universe experiences a period of exponential expansion driven by
the energy density of the vacuum. In addition, it is the
simplest model of space-time with a non-vanishing cosmological constant
$\Lambda$;  i.e. it is the simplest model of empty space-time but for the
energy of the vacuum. It shares with Minkowski space-time the property 
of being maximally symmetric, but on the other hand it has a non vanishing
curvature. Thus, it provides a tractable example of space-time in which the 
effects of curvature can be explicitly computed.    

Gravitational radiation is a subject of fundamental importance and current
interest in both astrophysics and cosmology. Its direct detection is now
a realistic challenging possibility within the current and expected
level of detector's sensitivity. Besides the weak field sources
(i.e. binary star systems, fissioning stars, oscillating and rotating
spheroids,...), neutron stars, collapsing supernovae, quasars and black
hole collisions are sources of intense gravitational radiation. In cosmology,
gravitational waves would arise during the several phase transitions
undergone by the early universe, as well as from string sources, and at the 
end of inflation \cite{Zeldovich}. The linearization procedure around 
flat Minkowski space-time is clearly not applicable in such strong
field problems. Perturbation techniques around curved backgrounds,
and in the presence of sources, are needed.

In this paper we consider the generation of gravitational waves by
generic sources which move in the de Sitter space-time. That is, we compute 
the
waves produced by  a source described by an energy-momentun tensor 
$T^{(1) \nu}_{\mu}$, which is covariantly conserved with respect to the
de Sitter metric. It turns out that the maximal symmetry of the
de Sitter space-time, and the use of conformal coordinates to  exploit its
conformal flatness, allow a discussion of gravitational radiation in this
space following analogous steps as in Minkowski space-time.
Of course, we consider weak enough sources for
the linear approximation to hold. Apart from this condition, the sources are
totally generic.

As one of the main results of our study, we obtain explicit closed 
formulae for
the spectrum -amplitude as a function of the frequency- of gravitational waves
in terms of a suitable spectral decomposition of the energy-momentum
tensor which is producing these waves. These formulae should  be interesting 
for primordial cosmology, since they relate the spectrum of primordial
gravity waves produced during the inflationary period to the energy-momentum
tensor of the possible sources existing during this period. This is a
classical  mechanism of production of primordial gravity waves, and it
is different to the quantum mechanical production of gravitons due to
the time variation of the de Sitter metric which has been studied before
\cite{Grishuk}, \cite{FordParker}.   

The paper is organized in a self-contained way as follows:
In section II  we consider a linearized gravitational perturbation around
the 4-D de Sitter background. Thus, we consider the pertubative expansion of 
the full Einstein equations -including the source and the metric 
perturbation- in orders of the perturbation $h_\mu^ \nu$. Due to the maximal 
symmetry of the de Sitter space-time, the  Einstein equations up to first 
order, can be consistently splitted into two equations: a  zero order equation
describing the 
de Sitter space being produced by the cosmological constant, and a first 
order equation which describes the production of gravity waves from a source.
Thanks to the maximal symmetry, the
covariant conservation of the source energy-momentum tensor $T^{(1)\nu}_{\mu}$ 
with respect to the de Sitter metric coincides with the 
integrability condition for the first order part of the Einstein equation. 
 
In section III we discuss the gauge invariance of the first order equation.
Again the maximal symmetry of the de Sitter space-time implies
the  gauge invariance of the first order part of the Einstein tensor
$G^{(1) \nu }_\mu$  under infinitesimal coordinate transformations. 
Then, in order to give the first order part of the Einstein equation
a definite form, we set a gauge fixing condition of the form
$D_\nu \psi_\mu^\nu = B_\mu$,
where $\psi_\mu^\nu$ is the trace reversed graviton field.  
The field $B_\mu$ is chosen as to
eliminate the non-diagonal terms in first derivatives from $G^{(1) \nu }_\mu$.
With this choice, we arrive to the simple decoupled system of  wave equations
(\ref{cojoeq}) governing the linearized gravitational field $\psi_\mu^\nu$ 
produced by the source $T^{(1)\nu}_{\mu}$.

In section IV we solve the homogeneous version of eq.(\ref{cojoeq}) which
describes free (sourceless) gravitational waves. 
The solutions turn out to be remarkably
simple as a consequence of our gauge choice (\ref{psifix}). They amount
to de Sitter minimally coupled massless scalar modes and Minkowski massless
modes \cite{TsamisWoodPL}. We also discuss the residual gauge invariance 
allowed by our gauge choice and use it to extract the two transverse
traceless physical polarizations of the graviton. We conclude this section
setting the conditions of validity for the linear approximation. 

Section V is devoted to the computation of the retarded Green's function
solving the graviton wave equation (\ref{cojoeq}).
The graviton retarded Green's function and also the Feynman propagator 
in de the Sitter space-time have been studied
in previous papers  \cite{AllenLN}, \cite{TsamisWoodComm}, following
a geometrical approach developed in \cite{AllenJacob}. In this paper 
we compute the  retarded Green's function most easily using QFT 
techniques. By  the way we also obtain a very simple expression for the
Schwinger commutator function of a minimally coupled massless scalar field.

Finally in section VI  we compute the linearized gravitational field produced
by a generic source with energy-momentum tensor $T^{(1)\nu}_{\mu}$.
We show that the  spectral decomposition of this  gravitational field in 
frequencies, can be easily expressed in terms of a suitable spectral 
decomposition of the energy momentum tensor $T^{(1)\nu}_{\mu}$. This spectral
decomposition is 
dictated by the form of the free gravitational wave modes. We also show that 
the linearized gravitational field  produced by  localized sources takes
the form of  free gravitational waves being radiated away from the source,
and whose amplitudes are easily related to $T^{(1)\nu}_{\mu}$.

We end up with the conclusions in section VII, followed by 
an appendix with the formulae for covariant
derivatives, curvature tensors and d'Alembertians for the de Sitter 
space-time which are needed in the paper.


\section{Gravitational perturbations in de Sitter space-time}
In this section we study the linearized gravitational perturbations around
the 4-D de Sitter background. Thus, we start from the metric

\begin{equation}
g_{\mu \nu} \equiv \gamma_{\mu \nu} + h_{\mu \nu} = \frac{1}{H^2 \eta^2} 
\left( \eta_{\mu \nu} + \phi_{\mu \nu} \right)
\label{met}
\end{equation}

\noindent
where $\eta_{\mu \nu} = {\rm diag}(-,+,+,+) $ is the Minkowski metric, 
$\gamma_{\mu \nu}$ is the de Sitter metric, and  $ h_{\mu \nu} = 
(H^2 \eta^2)^{-1} \phi_{\mu \nu} $ is a small perturbation, in the sense that
the components of the tensor density $\phi_{\mu \nu}$ are
much smaller than 1. 

As it is less involved to work with tensors rather than with tensor
densities, it is convenient to introduce the tensor field

\begin{equation}
h_\mu^\nu = \gamma^{\nu \rho} h_{\mu \rho}
\label{hache}
\end{equation}

\noindent
where $\gamma^{\nu \rho}$ is the inverse  de Sitter metric, and so

\begin{equation}
h_0^0 = -  \phi_{00} ~~~,~~~ h_i^0 = - h_0^i = -\phi_{0i}~~~,~~~ 
h_i^j = h_j^i = \phi_{ij}
\label{hachephi}
\end{equation}

It is also convenient to introduce the trace reversed graviton field

\begin{equation}
\psi_{\mu}^{\nu} =   h_\mu^\nu - \frac{1}{2}\, \delta_\mu^\nu\, h
\label{psi}
\end{equation}

\noindent
where $ h \equiv h_\lambda^\lambda $.

A remark about the conventions for the raising and lowering of tensor indices
is  here in order.  The indices of $h_\mu^\nu$ (or $\psi_\mu^\nu$) are raised
and lowered with the de Sitter metric,  $\gamma^{\mu \nu}$ denotes the 
inverse of $\gamma_{\mu \nu}$, $D_\mu$ denotes the covariant derivative with
respect to the de Sitter metric, and $ D^\nu = \gamma^{\nu \rho} D_\rho$. 
For the rest of the tensors, their indices are raised and lowered with the 
full metric and its inverse $g^{\mu \nu} $, which up to first order reads
$g^{\mu \nu} = \gamma^{\mu \nu} -  h^{\mu \nu} + 0(h^2)$. Notice however, 
that  when a tensor has a perturbative expansion, the former rule applies to 
the full tensor, and not to each of the terms in the expansion. Thus if we 
consider, for example,  the expansions for the full Ricci tensor up to first 
order in  $h_\mu^\nu$: $ R_\mu^\nu = R^{(0)\nu}_\mu + R^{(1)\nu}_\mu + 
\cdots$ and $R_{\mu \kappa} = R^{(0)}_{\mu \kappa}  + R^{(1)}_{\mu \kappa} +
 \cdots$,  we have $R_\mu^\nu = g^{\nu \rho}  R_{\mu \rho}$, which implies 
for the first order term
$ R^{(1)\nu}_\mu  = \gamma^{\nu \rho}  R^{(1)}_{\mu \rho} - h^{\nu \rho} 
R^{(0)\nu}_\mu$.

Now we consider the  Einstein equations
\begin{equation}
G_\mu^\nu \equiv R_\mu^\nu -  \frac{1}{2}\, \delta_\mu^\nu\ R =  
- 8 \pi {\mathcal{G}} \, T_\mu^\nu
\label{Einseq}
\end{equation}

\noindent
and deal with them in the following way: we develop the Einstein tensor 
$G_\mu^\nu$ up to first order in $h_\mu^\nu$

\begin{equation}
G_\mu^\nu = G^{(0)\nu}_\mu + G^{(1)\nu}_\mu + \cdots
\label{devG}
\end{equation}

\noindent
and for the second member we set 

\begin{equation}
T_\mu^\nu = T^{(0)\nu}_\mu + T^{(1)\nu}_\mu + \cdots
\label{devT}
\end{equation}

\noindent
where

\vspace{-0.2cm}

\begin{equation}
T^{(0)~\nu}_\mu =  - \Lambda \,  \delta_\mu^\nu 
\label{Tcero}
\end{equation}

\noindent
is the energy-momentum tensor producing the de Sitter background  with  
cosmological constant 
$\Lambda = \frac{3H^2}{8\pi {\mathcal{G}}}$, and $ T^{(1)\nu}_\mu $ is the 
energy-momentum tensor of a source.
This source is moving in the de Sitter background. So, $ T^{(1)\nu}_\mu $ 
is covariantly conserved with respect to the de Sitter metric, i.e.

 \begin{equation}
D_\nu T^{(1) \nu}_\mu =  0 
\label{covconsT}
\end{equation}
 
Thus, up to first order in $ h_\mu^\nu $,  the Einstein equations
(\ref{Einseq}) give the two equations

\begin{equation}
G^{(0)\nu}_\mu = -8\pi {\mathcal{G}}\, T^{(0)\nu}_\mu
\label{Einseqcero}
\end{equation}

\vspace{-0.2cm}

\noindent
and

\vspace{-0.2cm}

\begin{equation}
G^{(1)\nu}_\mu = -8\pi {\mathcal{G}} \,  T^{(1)\nu}_\mu
\label{Einsequno}
\end{equation}

Then,  equation (\ref{Einseqcero}) describes how the background is produced
by  the energy-momentum tensor $T^{(0)\nu}_\mu$,  and (\ref{Einsequno}) 
describes the production of a gravitational perturbation by the 
energy-momentum tensor $ T^{(1)\nu}_\mu$, which is to be considered small, 
i.e. of first order. This is exactly the point of view adopted when computing
gravitational radiation by astrophysical sources in Minkowski space-time 
\cite{Weinberg}, \cite{Miessner}, with the only difference that, for  
Minkowski space-time, $ T^{(0)\nu}_\mu$ vanishes. There is however a catch: 
we should  make sure that eq. (\ref{Einsequno}) is consistent, i.e.
integrable. Now, the integrability condition for the full Einstein equation 
(\ref{Einseq}) takes the form 

\begin{equation}
{\mathcal{D}}_\nu T_\mu^\nu = D_\nu T^{(0)\nu}_\mu + D_\nu T^{(1)\nu}_\mu +
\Gamma^{(1)\nu}_{\nu \rho}\, T^{(0)\rho}_\mu - 
\Gamma^{(1)\rho}_{\nu \mu}\, T^{(0)\nu}_\rho  + \cdots = 0
\label{devfullcovcons}
\end{equation}

\noindent
where ${\mathcal{D}}_\mu$ is the covariant derivative with respect to the
full metric $g_{\mu \nu}$. In eq. (\ref{devfullcovcons}),  we have 
developed the
metric connection up to first order as 

\begin{equation}
\Gamma^{\nu}_{\mu \lambda} = \Gamma^{(0)\nu}_{\mu  \lambda} + 
\Gamma^{(1)\nu}_{\mu  \lambda}+ \cdots
\label{devCris}
\end{equation} 

\noindent
with  $ \Gamma^{(0)\nu}_{\mu  \lambda}$ being the metric connection 
for the de Sitter metric given in (\ref{Criscero}), and 
$ \Gamma^{(1)\nu}_{\mu  \lambda}$ 
its first order correction

\begin{equation}\
\Gamma^{(1)\nu}_{\mu  \lambda} = \frac{1}{2} \left( D_\mu h_\lambda^\nu +  
D_\lambda  h_\mu^\nu -  D^\nu  h_\mu^\lambda \right) 
\label{Crisuno} 
\end{equation} 

Now the zero order part of eq. (\ref{devfullcovcons}), 
$D_\nu T^{(0)\nu}_\mu = 0 $, is obviously fulfilled, and the first 
order part yields the integrability condition 

\begin{equation}
D_\nu T^{(1)\nu}_\mu + \Gamma^{(1)\nu}_{\nu \rho}\,  T^{(0)\rho}_\mu - 
\Gamma^{(1)\rho}_{\nu \mu}\,  T^{(0)\nu}_\rho  = 0
\label{intconduno} 
\end{equation}

However, due to the maximally symmetric form (\ref{Tcero}) of 
$T^{(0)\nu}_\mu$, the terms proportional to $\Gamma^{(1)}$ identically 
cancel each other, and we 
are left with eq. (\ref{covconsT}) as integrability condition for 
eq. (\ref{Einsequno}). Thus,  equation (\ref{Einsequno}) can be consistently 
solved for any de Sitter covariantly conserved source $T^{(1)\nu}_\mu$, 
and it describes, as  we shall explain in this paper, the production of 
gravitational radiation by this source. In particular, for  
$T^{(1)\nu}_\mu = 0 $, we have 
the equation

\begin{equation}
G^{(1)\nu}_\mu = 0 
\label{Einsequnovac}
\end{equation}

\noindent
describing the propagation of free gravitational waves in the de Sitter 
space-time.

\section{Gravitational wave equation and choice of the gauge}

Once we have set equation (\ref{Einsequno}), our next step in order to 
solve it, will be to give it a definite form. This entails a combination of 
two things: to
obtain the expression of $G^{(1)\nu}_\mu$ in terms of $h_\mu^\nu $ and to fix 
the gauge. For the first,  we  start with the expression of the Ricci 
tensor in terms of the metric connection

\begin{equation}
R_{\mu \kappa} = \partial_\kappa 
\Gamma_{\mu \lambda}^\lambda - \partial_\lambda
\Gamma_{\mu \kappa}^\lambda
+ \Gamma_{\mu \lambda}^\rho \,  \Gamma_{\kappa \rho}^\lambda -
\Gamma_{\mu \kappa}^\rho \,  \Gamma_{\rho \lambda}^\lambda 
\label{Ricci}
\end{equation}

\noindent 
and expand it up to first order

\begin{equation}
R_{\mu \kappa} =  R_{\mu \kappa}^{(0)} +  R_{\mu \kappa}^{(1)}  + \cdots
\label{devRicci}
\end{equation}

\noindent
where $ R_{\mu \kappa}^{(0)} $ is given in (\ref{Riccicero}) and

\begin{equation}
R_{\mu \kappa}^{(1)} = \frac{1}{2} \left[ D^{\lambda} D_{\lambda} h_{\mu \kappa}
+ D_{\kappa} D_{\mu} h  - D_{\lambda} D_{\mu} h_\kappa^\lambda - 
D_{\lambda} D_{\kappa} h_\mu^\lambda \,   \right]
\label{Ricciuno}
\end{equation}

Then the l.h.s. of (\ref{Einsequno}) is given by 

\begin{equation}
G^{(1) \nu}_\mu  = R^{(1) \nu}_\mu  - \frac{1}{2}\, \delta_\mu^\nu \, R^{(1)} =
\gamma^{\nu \kappa} R^{(1)}_{\mu \kappa} - h^{\nu \kappa} R^{(0) \nu}_\mu 
- \frac{1}{2} \, \delta_\mu^\nu\, R^{(1)}
\label{Einstuno}
\end{equation}

\noindent
with $R^{(1)} = R^{(1) \lambda}_\lambda $
  
Next, we come to the gauge fixing. As a consequence of the covariance of the  
Einstein equations  (\ref{Einseq}), and the maximal symmetry of the de Sitter 
space-time, the equation for the gravitational perturbation  (\ref{Einsequno}) 
is gauge invariant under the transformations

\begin{equation}
h_{\mu \kappa}  \longrightarrow  h^{\prime}_{\mu \kappa} = h_{\mu \kappa} + 
D_\mu \xi_\kappa + D_\kappa \xi_\mu
\label{gaugeh}
\end{equation}

\noindent
where $\xi^\mu$ is an infinitesimal vector field to be considered of the same 
order as $h_\mu^\nu$, i.e. first order. Let us explain how this gauge 
invariance
arises. Under an infinitesimal coordinate transformation 

\begin{equation}
x^\mu   \longrightarrow   x^{\prime \mu}  =  x^\mu - \xi^\mu (x)   
\label{coordtransf}
\end{equation}

\noindent
the metric $g_{\mu \nu}$ transforms  as 

\begin{equation}
g_{\mu \nu}  \longrightarrow  g^{\prime}_{\mu \nu} =  
g_{\mu \nu} + {\mathcal{L}}_\xi \,  g_{\mu \nu}
= g_{\mu \nu}  + D_\mu \xi_\nu  + D_\nu  \xi_\mu  
\label{gcoordtransf}
\end{equation}

\noindent 
where  $\mathcal{L}_\xi$ is the Lie derivative for the vector field $\xi^\mu$, 
and we have omitted second order terms. Thus,  we can take the  point of 
view in which the  background metric $\gamma_{\mu \nu}$ remains invariant 
under the infinitesimal coordinate transformations (\ref{coordtransf}), 
while the 
perturbation $h_{\mu \nu}$ transforms as stated in (\ref{gaugeh}). Then, for the
tensors $G_\mu^\nu $ and $T_\mu^\nu $, the zero order parts remain invariant 
under (\ref{coordtransf}), while  the first order parts transform as

\begin{equation}
G^{(1) \nu}_\mu \longrightarrow   G^{\prime \,  (1) \nu}_\mu =  G^{(1) \nu}_\mu
+ {\mathcal{L}}_\xi \, G^{(0) \nu}_\mu
\label{Einscoordtransf}
\end{equation}

\noindent
and a similar expression for $ T^{(1) \nu}_\mu $. But now, the maximally 
symmetric forms $ G^{(0) \nu}_\mu \propto \delta_\mu^\nu$ and $ T^{(0) \nu}_\mu
\propto \delta_\mu^\nu$, imply ${\mathcal{L}}_\xi \,  G^{(0) \nu}_\mu = 0 $ 
and ${\mathcal{L}}_\xi \,  T^{(0) \nu}_\mu  = 0$. Thus,  $ G^{(1) \nu}_\mu$ is  
invariant under the infinitesimal coordinate transformations 
(\ref{coordtransf}),  as can also be directly checked by substitution of the 
gauge transformation (\ref{gaugeh}) in the expression (\ref{Einstuno}). 
Hence, eq. (\ref{Einsequno}) is a second order partial  differential equation 
for the unknown function $h_\mu^\nu$, with a gauge invariant source  
$T^{(1) \nu}_\mu$, which is gauge invariant under the transformation 
(\ref{gaugeh}). Therefore, in order to solve it, we need to set a gauge fixing
 condition 
on  $h_\mu^\nu$ (or $\psi_\mu^\nu$). Of  course,  if we were to solve only the  
homogeneous equation (\ref{Einsequnovac}),  the simplest procedure  would be to 
impose the gauge condition \cite{Grishuk}, \cite{FordParker}.

\begin{equation}
D_\nu \psi^\nu_\mu = 0 ~~~,
\label{deDonder}
\end{equation}

\noindent
and then use the residual gauge invariance allowed by (\ref{deDonder}), for 
going the so called  TT gauge \cite{Miessner}, 
by means of the additional condition 
$\psi_\mu^0 = 0$. However, this cannot be done in the presence of a 
non-vanishing source term $T^{(1) \nu}_\mu$. The situation here is  similar to 
the Maxwell equations with a source term: $\Box A_\mu - \partial_\mu 
(\partial_\nu A^\nu) = J_\mu$.  Once we have chosen the Lorentz gauge 
$\partial_\nu A^\nu = 0$,  we cannot impose in addition $A_0 = 0$, unless 
$J_0 = 0$. It is true that we  could set indeed the gauge condition 
(\ref{deDonder}), and then proceed to solve the resulting gauge fixed form of 
(\ref{Einsequno}). However, as we shall explain below, this introduces spurious 
(gauge) complications in the mode solutions to (\ref{Einsequno}). Thus, we shall
impose the  more general gauge fixing condition 

\begin{equation}
D_\nu \psi^\nu_\mu = B_\mu
\label{gaugefix}
\end{equation}

\noindent
and let the equations choose their favourite gauge fixing field $B_\mu$ 
instead. (Notice that eq. (\ref{gaugefix}) implies that the gauge fixing field 
$B_\mu$ is to be considered of first order in the perturbation expansion).  

Imposing the gauge condition (\ref{gaugefix}), the first order part of the 
Ricci tensor (\ref{Ricciuno})can be written as 

\begin{equation}
R_{\mu \kappa}^{(1)} = \frac{1}{2} \left[\, D^{\lambda} D_{\lambda}\, 
 h_{\mu \kappa} + 2\,h^{\lambda \sigma} R^{(0)}_{\sigma \mu \kappa \lambda} 
+ h_{\mu}^{\sigma} R^{(0)}_{\sigma  \kappa } 
+ h_{\kappa}^{\sigma} R^{(0)}_{\sigma  \kappa } 
- D_{\kappa} B_{\mu} -   D_{\mu} B_{\kappa}\,  \right]
\label{RicciunoB}
\end{equation}

\noindent
Then from (\ref{RicciunoB}) and (\ref{Einstuno}), the first order part of the 
Einstein tensor is 

\begin{eqnarray}
G^{(1) \nu}_\mu     & = &  \frac{1}{2}\, D^{\lambda} D_{\lambda}\, \psi_\mu^\nu 
+ R_{\lambda \sigma \mu}^{(0) \nu}\, h^{\lambda \sigma} + \frac{1}{2}\, 
\delta_\mu^\nu \, R_{\lambda \sigma }^{(0) }\,  h^{\lambda \sigma} \nonumber \\
& & + \frac{1}{2} \left[ - D^\nu B_\mu - D_\mu B^\nu + \delta_\mu^\nu\,  
D_\lambda B^\lambda  \right]    
\label{EinstunoB}
\end{eqnarray}

\noindent
Now according to (\ref{divercovpsi}), the gauge fixing condition 
(\ref{gaugefix}) reads

\begin{equation}
\partial_\lambda  \psi_\mu^\lambda  -  \frac{4}{\eta}\,  \psi_\mu^0 +
\frac{1}{\eta}\,  \delta_\mu^0 \, \psi = B_\mu
\label{devgaugefix}
\end{equation}

\noindent
This allows us to rewrite the two terms in the second line of the tensor 
d'Alembertian (\ref{tensordal}) as 

\begin{eqnarray}
\delta_\mu^0 \,  \eta^{\nu \alpha } \,  \partial_\beta \psi_\alpha^\beta 
- \delta_0^\nu \, \partial_\beta \psi_\mu^\beta & = & - \frac{4}{\eta} 
\left(\delta_\mu^0 \, \psi_0^\nu + \delta_0^\nu \, \psi_\mu^0 \right) 
+ \frac{2}{\eta}\, \delta_\mu^0 \, \delta_0^\nu \, \psi \nonumber \\
& & + \delta_\mu^0 \, \eta^{\nu \alpha}\, B_\alpha - \delta_0^\nu \,  B_\mu 
\label{simpgaugetensordal}
\end{eqnarray}

So, for a tensor field  $ \psi_\mu^\nu $ that satisfies the gauge fixing 
condition (\ref{devgaugefix}), the tensor d'Alembertian can be recast as

\begin{eqnarray}
\frac{1}{H^2 \eta^2} \, D^{\lambda} D_{\lambda}\, \psi_\mu^\nu & = & 
\Box \, \psi_\mu^\nu  + \frac{2}{\eta} \left[\, \partial_\eta \psi_\mu^\nu 
+ \partial_\mu \psi_0^\nu - \eta^{\nu \kappa} \, \partial_\kappa  \psi_\mu^0 \, 
\right] \nonumber \\
& & + \frac{2}{\eta^2} \left[\, \psi_\mu^\nu + \delta_\mu^0 \, \delta_0^\nu \, 
\psi - 2 \, \delta_\mu^0 \, \psi_0^\nu - 2 \, \delta_0^\nu \,  \psi_\mu^0  - 
\delta_\mu^\nu \, \psi_0^0 \, \right] \nonumber \\
& & +\frac{2}{\eta} \left[\,  \delta_\mu^0 \, \eta^{\nu \alpha}\, B_\alpha -  
\delta_0^\nu\, B_\mu \,  \right]
\label{gaugetensordal}
\end{eqnarray}   

\noindent
In addition, from eqs. (\ref{Riemanncero}), (\ref{Riccicero}) and  (\ref{psi}),
we have 

\begin{equation}
\frac{1}{H^2 \eta^2} \,  R^{(0) \nu}_{\lambda \sigma \mu} \,  h^{\lambda \sigma}
  = - \frac{1}{\eta^2} \left(\psi_\mu^\nu  + \frac{1}{2} \, \delta_\mu^\nu \,  
\psi  \right)
\label{psiRiemanncero}
\end{equation}

\noindent
and
\vspace{-0.2cm}

\begin{equation}
\frac{1}{H^2 \eta^2} \, R^{(0) }_{\lambda \sigma } \, h^{\lambda \sigma}  =
 \frac{3}{\eta^2} \,  \psi  
\label{psiRiccicero}
\end{equation}

\noindent
Also, using the  expressions (\ref{covdercov}) and (\ref{covdercontra}), 
the second line in eq. (\ref{EinstunoB}) takes the form
\begin{eqnarray}
\frac{1}{H^2 \eta^2}\, \left[ - D^\nu B_\mu - D_\mu B^\nu + \delta_\mu^\nu \,  
D_\lambda B^\lambda  \right] & = & - \eta^{\nu \alpha }\, \partial_\mu B_\alpha 
-  \eta^{\nu \alpha }\, \partial_\alpha  B_\mu + \delta_\mu^\nu \, 
\eta^{ \alpha \beta} \,  \partial_\alpha  B_\beta \nonumber \\
& & + \frac{2}{ \eta} \, \left( \delta_0^\nu \,  B_\mu - \delta_\mu^0 \,  
\eta^{\nu \alpha }\,  B_\alpha \right)   
\label{explicitbes}
\end{eqnarray}

\noindent
Finally, adding up the expressions (\ref{gaugetensordal}), 
(\ref{psiRiemanncero}), (\ref{psiRiccicero}) and  (\ref{explicitbes}),
we obtain 
 
\begin{eqnarray}
W_{\mu}^{\nu}  \equiv  \frac{2}{H^2\eta^2} \,  G^{(1) \nu}_\mu     & = &     
\Box \,  \psi_\mu^\nu + \frac{2}{\eta} \,  \partial_\eta  \psi_\mu^\nu 
+  \frac{2}{\eta} \left( \partial_\mu  \psi_0^\nu - \eta^{\nu \kappa} \,
 \partial_\kappa  \psi_\mu^0  \right)  \nonumber  \\ 
&  & +  \frac{2}{\eta^2} \left[  \left(\delta_\mu^\nu + \delta_\mu^0 \,  
\delta_0^\nu  \right) \psi - \delta_\mu^\nu \, \psi_0^0  - 2\, \delta_\mu^0 \, 
\psi_0^\nu  - 2\,  \delta_0^\nu \, \psi_\mu^0 \, \right] \nonumber \\
&  & - \eta^{\nu \kappa} \, \partial_\mu B_\kappa  - \eta^{\nu \kappa} 
\partial_\kappa B_\mu + \delta_\mu^\nu \,  \eta^{\lambda  \kappa} \, 
\partial_\lambda B_\kappa  
\label{cojoEinstunoB}
\end{eqnarray}

This expression suggests a suitable choice for $B_\mu$, namely 

\begin{equation}
B_\mu = 2H\, u_\nu \psi_\mu^\nu  
\label{Bmu}
\end{equation}

\noindent
where $u^\nu = H\eta \, \delta_0^\nu $ is a unit time-like vector in the 
background. With this choice, the non-diagonal terms in first derivatives in the
first line of eq.  (\ref{cojoEinstunoB}),  cancel against the first two terms of 
the third line. Moreover, with the choice (\ref{Bmu}), the gauge fixing 
condition (\ref{gaugefix}) takes the form

\begin{equation}
D_\lambda \psi_\mu^\lambda - 2 H \, u_\lambda \psi_\mu^\lambda \equiv 
\partial_\lambda \psi_\mu^\lambda - \frac{2}{\eta} \, \psi_\mu^0 + \frac{1}{\eta}
\, \delta_\mu^0 \,  \psi = 0
\label{psifix}
\end{equation}

Then, using eqs. (\ref{Bmu})  and (\ref{psifix}), we also  have 

\begin{equation}
\eta^{\lambda  \kappa} \partial_\lambda B_\kappa = \frac{2}{\eta^2} 
\left( \psi_0^0 - \psi \right)
\label{furthercan}
\end{equation}

\noindent 
which produces a further cancelation of two terms in  (\ref{cojoEinstunoB}). 
Therefore, we arrive at the following final gauge fixed form for the 
gravitational wave equations with a generic source term

\begin{equation}
\Box \,  \psi_\mu^\nu + \frac{2}{\eta} \,  \partial_\eta   \psi_\mu^\nu 
+ \frac{2}{\eta^2} \left[\, \delta_\mu^0 \,  \delta_0^\nu \,   \psi - 
\delta_\mu^0 \, \psi_0^\nu - \delta_0^\nu \, \psi_\mu^0 \,  \right]  =
- \frac{16 \pi \mathcal{G}}{H^2\eta^2} \,  T_\mu^{(1) \nu}
\label{cojoeq}
\end{equation}

Notice that this wave equation -which we have obtained by imposing the gauge 
condition (\ref{psifix})- is much simpler than the one  we would  have  obtained 
imposing (\ref{deDonder}), i.e. setting $ B_\mu = 0 $ in eq. 
(\ref{cojoEinstunoB}). The reader can also directly check that the solutions to 
(\ref{cojoeq}), satisfy the gauge condition (\ref{deDonder}) 
provided $T_\mu^{(1) \nu}$  is de Sitter covariantly conserved. 

Once we have obtained the graviton wave equation (\ref{cojoeq}), our next task 
will be to solve it for a generic $T_\mu^{(1) \nu}$. For this purpose, we need 
the retarded Green's function for the differential operator in  the left hand 
side of eq. (\ref{cojoeq}). This Green's function will be computed in section IV
as an appropriate superposition of solutions to the homogeneous version of
eq. (\ref{cojoeq}). Thus, we discuss first the solutions of the homogeneous 
equation, which  are interesting in their own right,  since they represent the 
free gravitational waves in  de Sitter space-time.


\section{Free gravitational waves in de Sitter space-time}
In this section we set the source $T_\mu^{(1) \nu} = 0 $, and solve the 
homogeneous version of (\ref{cojoeq}).

\begin{equation}
\Box\, \psi_\mu^\nu + \frac{2}{\eta}\, \partial_\eta  \psi_\mu^\nu 
+ \frac{2}{\eta^2} \left[\,\delta_\mu^0 \, \delta_0^\nu \,  \psi - \delta_\mu^0
\, \psi_0^\nu - \delta_0^\nu\, \psi_\mu^0\, \right]  = 0
\label{homocojoeq}
\end{equation} 

We shall see that in the gauge (\ref{psifix}) we have a very simple basis of 
mode solutions to the free gravitational wave equations (\ref{homocojoeq}). 
To describe the solutions of (\ref{homocojoeq}) it is convenient to introduce 
the tensor density

\begin{equation}
\chi_{\mu \nu} = H^2 \eta^2\, \gamma_{\nu \rho}\, \psi_\mu^\rho = 
\eta_{\nu \rho}\, \psi_\mu^\rho
\label{chi}
\end{equation}

\noindent
whose components are related to those of  $\psi_\mu^\nu$ by 

\begin{equation} 
\chi_{0 0} = -  \psi_0^0 ~~~,~~~ \chi_{0 i} = -  \psi_i^0 = \psi_0^i  ~~~,~~~ 
\chi_{i j} =  \psi_i^j = \psi_j^i
\label{chipsi}
\end{equation}

In terms of $\chi_{\mu \nu}$ the metric perturbation $\phi_{\mu \nu}$ is

\begin{equation}
\phi_{\mu \nu} = \chi_{\mu \nu} - \frac{1}{2}\,  \eta_{\mu \nu}\,  
\eta^{\alpha  \beta}\,  \chi_{\alpha  \beta}
\label{chiphi}
\end{equation} 

\noindent
We shall also use the notation

\begin{equation}
\hat{\chi} = \chi_{11} + \chi_{22} + \chi_{33} ~~~,~~~ \tilde{\chi} = \chi_{00} 
+ \hat{\chi}
\label{gorrochi}
\end{equation}

Then,  splitting  eq. (\ref{homocojoeq}) into its time-time, time-space, and 
space-space components, we obtain

\begin{equation}
\Box \,  \chi_{0 0} + \frac{2}{\eta}\, \partial_\eta \chi_{0 0}  - 
\frac{2}{\eta^2}\,   \tilde{\chi} = 0
\label{eqchitt}
\end{equation}

\vspace{-0.4cm}

\begin{equation}
\Box \, \chi_{0 i} + \frac{2}{\eta}\,\partial_\eta  \chi_{0 i} - 
\frac{2}{\eta^2}\, \chi_{0 i}  = 0
\label{eqchits}
\end{equation}

\vspace{-0.4cm}

\begin{equation}
\left( \Box \,  + \frac{2}{\eta} \, \partial_\eta \right) \chi_{ij} = 0
\label{eqchiss}
\end{equation}

From eqs. (\ref{scaldal}) and (\ref{eqchiss}), we see that the space components 
$\chi_{ij}$ behave as minimally coupled massless scalar fields in de Sitter 
space-time.  As it is known \cite{FordParker}, the two physical TT components of
a free gravitational wave in de Sitter space-time -which count among the 
$\chi_{ij}$-  behave as minimally coupled massless scalar fields. As we 
mentioned above, this result can be easily obtained in the absence of sources 
by sequentially imposing the two gauge conditions $D_\nu \psi_\mu^\nu = 0$ and 
$\psi_\mu^0 = 0$. Then, one  good property of the gauge condition (\ref{psifix})
 - which can be applied in the presence of sources-  is that it alone is 
sufficient to capture the minimally coupled massless scalar field behaviour 
(\ref{eqchiss}). In addition, the two other equations (\ref{eqchits}) and  
(\ref{eqchitt}) are very easy to solve. In fact, eq. (\ref{eqchits}) can be 
rewritten as 

\begin{equation}
\Box \, \left(\frac{1}{\eta}\, \chi_{0 i}  \right) = 0
\label{etaeqchits}
\end{equation}

\noindent
Moreover, adding eq. (\ref{eqchitt}) and the trace of eq. (\ref{eqchiss}) we 
also have 

\begin{equation}
\Box \,  \left(\frac{1}{\eta}\,\tilde{\chi}   \right) = 0
\label{eqchitilde}
\end{equation}

\noindent
Thus, the time-space components $\chi_{0i}$ and the combination $\tilde{\chi}$
behave as free massless Minkowski fields rescaled by a factor $\eta$.

Now it is very easy to write down the general solution for $\chi_{\mu \nu}$ as 
a superposition of plane waves

\begin{equation}
\chi_{\mu \nu} \left(\eta,\vec{x}\right) = \int d^3 \vec{k}\, \left[\, 
f_{\mu \nu} \left( \eta ;\vec{k} \right)\, \exp \left(i\vec{k}\vec{x}\, \right) 
+ f_{\mu \nu}^{\ast} \left(\eta ;\vec{k} \right) \, \exp 
\left( -  i\vec{k}\vec{x}\, \right) \,  \right]
\label{expwave}
\end{equation} 

\noindent
For $\chi_{0 i}$ and  $\tilde{\chi}$ we obviously have

\begin{equation}
\chi_{0 i}\left(\eta,\vec{x}\right) = \int d^3 \vec{k}\,  \left[\,   
e_{0 i}\left(\vec{k}\right) \, \eta \, 
\exp \left(-i\omega \eta + i\vec{k}\vec{x}\, \right)  
+  e_{0 i}^{\ast} \left( \vec{k} \right) \, \eta \,
\exp \left(i\omega \eta - i\vec{k}\vec{x} \, \right)\,   \right]
\label{expwavets}
\end{equation}

\noindent
and

\begin{equation}
\tilde{\chi} \left(\eta,\vec{x}\right) = \int d^3 \vec{k}\,  \left[\, 
\tilde{e}\left(\vec{k}\right) \, \eta\, 
\exp \left(-i\omega \eta + i\vec{k}\vec{x}\, \right)   
+ \tilde{e}^{\ast} \left( \vec{k}\right) \, \eta \,   
\exp \left(i\omega \eta - i\vec{k}\vec{x} \, \right)\,   \right]
\label{expwavetilde}
\end{equation} 

\noindent
where $\omega = |\vec{k}|$.

In addition, the ordinary differential equation for the modes 
$f_{ij}(\eta ;\vec{k})$ can be solved in terms of Hankel 
functions of index 3/2  \cite{GradRyz}. So, we have

\begin{equation}
\chi_{ij} \left( \eta,\vec{x} \right)  =  \int d^3\vec{k} \, \left[\,  
e_{ij}\left( \vec{k} \right) \left(\eta - \frac{i}{\omega} \right) \exp 
\left(-i\omega \eta + i\vec{k}\vec{x}\, \right)
+ e_{ij}^{\ast} \left( \vec{k} \right) \left(\eta + \frac{i}{\omega} \right)
\exp \left(i\omega \eta - i\vec{k}\vec{x} \, \right)\,   \right] 
\label{expwavess}
\end{equation}

\noindent
And from equation (\ref{gorrochi}) and (\ref{expwavetilde}) 

\begin{eqnarray}
\chi_{00} \left( \eta,\vec{x} \right)  =  \int d^3\vec{k} \, \left\{ \right. 
& & \left[\, \tilde{e}\left( \vec{k} \right)  \eta - \hat{e}\left( 
\vec{k} \right) \left(\eta - \frac{i}{\omega} \right) \right] 
\exp \left(-i\omega \eta + i\vec{k}\vec{x}\, \right)  \nonumber \\
& & + \left[ \,  \tilde{e}^{\ast} 
\left( \vec{k} \right) \eta - \hat{e}^{\ast} \left( \vec{k} \right)  
\left(\eta + \frac{i}{\omega} \right) \right] \exp 
\left(i\omega \eta - i\vec{k}\vec{x} \, \right)   \left.  \right\} 
\label{expwavett}
\end{eqnarray}

\noindent
where we have defined $\hat{e} ( \vec{k} )  = \delta_{ij}  e_{ij} ( \vec{k} )$.

Eqs. (\ref{expwavets})-(\ref{expwavess}) provide the general solution for a free
gravity wave in de Sitter space-time in the gauge (\ref{psifix}). As we shall 
see in the next section, the simplicity of these solutions will allow us a 
direct derivation of the graviton retarded Green's function. In this respect, 
it is enlightening to point out, that if we had imposed $D_\nu \psi_\mu^\nu = 0
\, $ as our gauge fixing condition, we would have obtained a much more
difficult coupled system of partial differential equations for $\chi_{\mu \nu}$ 
than eqs. (\ref{eqchitt})-(\ref{eqchiss}). In particular, the solutions to 
those equations involve Bessel functions of index $\nu = \sqrt{33}/2$.    
   
In the rest of this section, we show how the physical degrees of freedom for the
gravitational waves are contained in expressions (\ref{expwavets}),
(\ref{expwavetilde}), and (\ref{expwavess}). Due to the linearity, it is enough 
to consider just one mode with wave vector $\vec{k}$. To begin with, the gauge 
fixing condition (\ref{psifix}) reduces the ten polarizations that we have in 
$\chi_{\mu \nu}$  down to six. More precisely:  for the field  $\chi_{\mu \nu}$,
the gauge fixing condition  (\ref{psifix})
reads 

\begin{equation}
- \partial_\eta \chi_{00} + \partial_i  \chi_{0i} + \frac{1}{\eta} \, 
\tilde{\chi} = 0
\label{chifixt}
\end{equation}

\vspace{-0.4 cm}

\begin{equation}
- \partial_\eta \chi_{0i} + \partial_j  \chi_{ji} + \frac{2}{\eta} \, 
\chi_{0i} = 0
\label{chifixs}
\end{equation}

\noindent
which translated to the modes $f_{\mu \nu}(\eta ;\vec{k} )\, \exp 
\left( i\vec{k}\vec{x} \right) $ gives the following  constraints
among  polarizations

\begin{equation}
n_ie_{0i} + \tilde{e} - \hat{e} = 0
\label{efixt}
\end{equation}

\vspace{-0.2cm}

\noindent
and

\vspace{-0.2cm}

\begin{equation}
e_{0i} + n_j e_{ji} = 0
\label{efixs}
\end{equation}

\noindent 
where $n_i = k_i/\omega  $ is the unit wave vector. 

Next, the residual gauge invariance in the gauge fixing (\ref{psifix}) allows 
to reduce the six independent polarizations $e_{ij}$ to the two physical 
polarizations. Under the  gauge transformation (\ref{gaugeh}), the trace 
reversed graviton field $ \psi_\mu^\nu \, $ transforms as

\begin{equation}
\psi_\mu^\nu \longrightarrow  \psi_\mu^{ \prime \nu } = \psi_\mu^\nu  - 
D_{\mu} \xi^{\nu} -  D^{\nu} \xi_{\mu} + \delta_\mu^\nu \, 
D_{\lambda} \xi^{\lambda}
\label{gaugepsi}
\end{equation}

\noindent
Then, the invariance of the gauge condition  (\ref{psifix}) under the  
transformation (\ref{gaugepsi}) requires $\xi_\mu $ to be a  solution of 
the equation

\begin{equation}
D^{\lambda}D_{\lambda} \, \xi_\mu - \xi^\lambda R^{(0)}_{\lambda \mu } 
+ 2H \left(  \, u_\mu \,  D_{\lambda} \xi^{\lambda} - u_\lambda \,  
D_{\mu} \xi^{\lambda} -  u_\lambda \,  D^{\lambda} \xi_\mu \right) = 0
\label{eqxi}
\end{equation}

\noindent
From  eqs. (\ref{covdercov})-(\ref{divercov}),(\ref{Riccicero}), and 
(\ref{vectordal}) it reduces to 

\begin{equation}
\Box \,\xi_\mu - \frac{2}{\eta} \, \partial_\eta \xi_\mu   + \frac{2}{\eta^2}\, 
\left(\xi_\mu - \delta_\mu^0 \, \xi_0 \right) = 0 
\label{eqxisimp}
\end{equation}

\noindent
which yields the simple decoupled equations

\begin{equation}
\Box \left( \frac{1}{\eta} \xi^0  \right) = 0
\label{eqxit}
\end{equation}

\vspace{-0.2cm}

\noindent
and

\vspace{-0.2cm}

\begin{equation}
\left( \Box + \frac{2}{\eta} \, \partial_\eta   \right) \left( 
\frac{1}{\eta} \xi^i  \right) = 0
\label{eqxis}
\end{equation}


The solutions to eqs.(\ref{eqxit}) and (\ref{eqxis}) decompose in modes in the 
form

\begin{equation}
\xi^0 \left( \eta,\vec{x} \right) = \int d^3 \vec{k}\, \left[\, i\, 
\varepsilon_0 \left(\vec{k} \right) \,  \eta \,  
\exp \left(-i\omega \eta + i\vec{k}\vec{x}\, \right) 
- i\,  \varepsilon^{\ast}_0 \left( \vec{k} \right) \,  \eta \,  
\exp \left(i\omega \eta - i\vec{k}\vec{x} \, \right)\,   \right]
\label{xit}
\end{equation}

\noindent
and

\begin{equation}
\xi^i \left( \eta,\vec{x} \right)  = \int d^3 \vec{k}\, \left[\, i\, 
\varepsilon_i \left( \vec{k} \right) \left(\eta - \frac{i}{\omega} \right) \exp 
\left(-i\omega \eta + i\vec{k}\vec{x}\, \right) - i \,  \varepsilon^{\ast}_i 
\left( \vec{k} \right)   \left(\eta + \frac{i}{\omega} \right) 
\exp \left(i\omega \eta - i\vec{k}\vec{x} \, \right)\,   \right]
\label{xis}
\end{equation}

On the other hand, from the transformation law (\ref{gaugepsi}) we find

\begin{equation}
\delta_\xi \,  \chi_{0i} = \partial_i \xi^0 -\partial_\eta \xi^i
\label{deltachits}
\end{equation}

\begin{equation}
\delta_\xi \, \chi_{ij} = - \partial_i \xi^j  - \partial_j \xi^i + \delta_{ij} 
\left( \partial_\eta \xi^0 + \partial_l \xi^l - \frac{2}{\eta}\, \xi^0 \right)
\label{deltachiss}
\end{equation}

\begin{equation}
\delta_\xi \,  \tilde{\chi} = 4 \, \partial_\eta \xi^0 - \frac{4}{\eta}\, \xi^0
\label{deltachitilde}
\end{equation}

\noindent
Then, from eqs. (\ref{deltachits})-(\ref{deltachitilde}), the residual gauge 
transformations given by (\ref{xit})-(\ref{xis}), amount to the following 
transformation laws for polarizations

\begin{eqnarray}
\tilde{e}  \longrightarrow \tilde{e}^{\prime} & = & \tilde{e} + 4\omega 
\varepsilon_0 \nonumber \\ 
e_{0i}   \longrightarrow   e_{0i}^{\prime}  & = & e_{0i} - k_i \varepsilon_0 - 
\omega \varepsilon_i \nonumber \\
e_{ij}   \longrightarrow   e_{ij}^{\prime}  & = & e_{ij} + k_i \varepsilon_j +  
k_j \varepsilon_i + \delta_{ij} \left( \omega \varepsilon_0 - \vec{k} 
\vec{\varepsilon} \right)    
\label{poltransf}
\end{eqnarray}

Thus, it is possible to eliminate the polarizations $e_{0i}$ and $\tilde{e}$ by 
suitably choosing the parameters $\varepsilon_\mu$. Therefore, we can set the 
supplementary  gauge conditions

\begin{equation}
\chi_{0i} = \tilde{\chi} = 0
\label{supgauge}
\end{equation}  

\noindent
Of course, the gauge transformations (\ref{poltransf}) are compatible with the 
constraints (\ref{efixt}) and  (\ref{efixs}), and by replacing the 
supplementary conditions $e_{0i} = \tilde{e} = 0$ in eqs. 
(\ref{efixt}) and  (\ref{efixs}) we find

\begin{equation}
\hat{e} = n_j e_{ji} = 0
\label{TTcond}
\end{equation}

\noindent
This leave us with the traceless transverse (TT) graviton physical polarizations.
Notice that since $\hat{e} = 0$, eq. (\ref{expwavett}) implies  $\chi_{00} =0$.
Therefore, $\phi_{\mu \nu}$  coincides with $\chi_{\mu \nu}$, and the TT 
conditions are satisfied in  this gauge by the metric perturbation itself.

We conclude this section with a comment about the validity of the linear 
approximation  we apply. Let us consider just one physical mode for  
$\phi_{\mu \nu}$,  with amplitude $A$ and  wave vector $\vec{k}$.  

\begin{equation}
A  \left(\eta - \frac{i}{\omega} \right) \exp \left(-i\omega \eta 
+ i\vec{k}\vec{x}\, \right) + A^{\ast}  \left(\eta + \frac{i}{\omega} \right) 
\exp \left(i\omega \eta - i\vec{k}\vec{x}\, \right) 
\label{mode}
\end{equation}

\noindent
As stated at the beginning of section II, the validity of the linear 
approximation requires  $|\phi_{\mu \nu}| \ll 1 $. Thus, according to the 
functional form of the mode (\ref{mode}), the amplitude $A$ must satisfy 
the conditions

\begin{equation}
|A| \ll \frac{1}{|\eta|} ~~~,~~~  |A| \ll \omega
\label{condA}
\end{equation}

\noindent
that is, the linear approximation holds for high frequencies and small conformal
time. Let us remind that the cosmic region for  the  de Sitter space-time which 
describes  an exponentially expanding space-time, corresponds to
$ -1/H < \eta < 0$. Thus, the linear approximation holds  throughout this 
region provided that 

\begin{equation}
 A \ll \max \left\{ H,\, \omega \right\}   
\label{condAcosm}
\end{equation}


\section{Retarded Green's function for the minimally coupled massless scalar 
field and the graviton}

We come back now to the inhomogeneous equation (\ref{cojoeq}).
Splitting (\ref{cojoeq}) we have 

\begin{equation}
\Box \, \chi_{0 0} + \frac{2}{\eta}\, \partial_\eta \chi_{0 0}  
- \frac{2}{\eta^2}\, \tilde{\chi} =
- 16\, \pi \, { \mathcal{G}} \,  T_{00}^{(1)}
\label{inhomchitt}
\end{equation}

\vspace{-0.4cm}

\begin{equation}
\Box \, \chi_{0 i} + \frac{2}{\eta}\,\partial_\eta  \chi_{0 i} - 
\frac{2}{\eta^2}\, \chi_{0 i}  =  - 16\, \pi \, {\mathcal{G}} \,  T_{0i}^{(1)}
\label{inhomchits}
\end{equation}

\vspace{-0.4cm}

\begin{equation}
\left( \Box \,  + \frac{2}{\eta} \, \partial_\eta \right) \chi_{ij} = 
- 16\, \pi \, {\mathcal{G}}  \,  T_{ij}^{(1)}
\label{inhomchiss}
\end{equation}

\noindent
As in previous section, eq. (\ref{inhomchits}) can be rewritten as

\begin{equation}
\Box \, \left(\frac{1}{\eta}\, \chi_{0 i}  \right) =  - \frac{16\, \pi \, 
\mathcal{G}}{\eta}\, T_{0i}^{(1)} 
\label{inhometachits}
\end{equation}

\noindent
and adding up eq. (\ref{inhomchitt}) with the trace of (\ref{inhomchiss}) we 
also have

\begin{equation}
\Box \, \left(\frac{1}{\eta}\, \tilde{\chi}  \right) =  
- \frac{16\, \pi \, \mathcal{G}}{\eta} \, \tilde{T}^{(1)} 
\label{inhometachitilde} 
\end{equation}

\noindent
Here

\begin{equation}
\hat{T}^{(1)} = T^{(1)}_{11} + T^{(1)}_{22} + T^{(1)}_{33} ~~~,~~~ \tilde{T} = 
T^{(1)}_{00} + \hat{T}^{(1)}
\label{gorroT}
\end{equation}

\noindent
Thus, the retarded solution to eqs. (\ref{inhometachits}) and 
(\ref{inhometachitilde}) can be obtained using the well known
Minkowski massless retarded Green's function

\begin{equation}
G_R^{(M)}(x,x^{\prime}) = \frac{1}{4\, \pi \, |\vec{x} - \vec{x}^{\prime}|}\, 
\delta \left( \eta - \eta^{\prime} - |\vec{x} - \vec{x}^{\prime} |\, \right)
\label{GreenMink}
\end{equation}

\noindent
which has support on the past light cone and satisfies

\begin{equation}
\Box \, G_R^{(M)}(x,x^{\prime}) = - \delta^{(4)}(x - x^{\prime}) 
\label{eqGreenMink}
\end{equation}

On the other hand, in order to solve (\ref{inhomchiss}) we need the retarded 
Green's function for the scalar d'Alembertian in de Sitter space-time, 
i.e. we need the retarded solution to

\begin{equation}
D^\lambda_\lambda \,  G_R(x,x^{\prime}) = - H^4 \eta^4 \, 
\delta^{(4)}(x - x^{\prime})  
\label{eqscal}
\end{equation}

\noindent
Here  $D^\lambda_\lambda$ is the scalar D'alambertian in de Sitter space-time
given in (\ref{scaldal}). Thus, eq. (\ref{eqscal}) takes the form

\begin{equation} 
\left( \Box +\frac{2}{\eta} \,  \partial_{\eta} \right) G_R(x,x^{\prime}) = 
-  H^2 \eta^2 \, \delta^{(4)}(x - x^{\prime}) 
\label{simpeqscal}
\end{equation}

This retarded Green's function $ G_R(x,x^{\prime})$ could be obtained
as in \cite{TsamisWoodComm}, using the geometric techniques developed in   
\cite{AllenJacob}. However, although the retarded Green's function is a 
purely classical object, it can be most easily obtained using QFT techniques. 
Thus, we consider a scalar field 

\begin{equation} 
\phi (\eta, \vec{x}) = \int d^3 \vec{k}\, \left[\, u_{\vec{k}}(\eta, \vec{x})\, 
a(\vec{k}) + u^{\ast}_{\vec{k}}(\eta, \vec{x})\, a^{\dagger} (\vec{k}) \, \right]
\label{scalphi}
\end{equation}

\noindent
obeying the homogeneous equation 

\begin{equation} 
\left( \Box +\frac{2}{\eta} \,  \partial_{\eta} \right) \phi = 0    
\label{eqscalphi}
\end{equation}

\noindent
i.e. $ \phi (\eta, \vec{x})  $ is a minimally coupled scalar field in de 
Sitter space-time. Then, from the previous section, 
the modes $u_{\vec{k}}\, (\eta, \vec{x}) $ take the form

\begin{equation} 
u_{\vec{k}}(\eta, \vec{x}) = \frac{H}{(2\pi)^{3/2}\, \sqrt{2\omega}}
\left(\eta - \frac{i}{\omega} \right) \exp 
\left(-i\omega \eta + i\vec{k}\vec{x}\, \right)     
\label{scalmode} 
\end{equation}

\noindent
These modes satisfy the normalization

\begin{equation} 
\left(  u_{\vec{k}^{\prime}}\, , \,   u_{\vec{k}} \right) = 
\delta (\vec{k} - \vec{k}^{\prime} )   
\label{norm} 
\end{equation}

\noindent
with respect to the scalar product


\begin{equation} 
\left( \phi_2 \, , \, \phi_1  \right) =  \frac{i}{H^2 \eta^2} \int d^3 \vec{x} 
\, \, \phi_2^{\ast}(\eta, \vec{x}) 
\stackrel{\leftrightarrow}{\partial_\eta} \phi_1(\eta, \vec{x}) 
\label{scalproduct} 
\end{equation}

Now, if we canonically quantize the scalar field $\phi$ according to

\begin{equation} 
\left[ a(\vec{k}) \, , \,  a^{\dagger} (\vec{k^{\prime}}) \right] = 
\delta (\vec{k} - \vec{k}^{\prime} )    
\label{commut} 
\end{equation}

\noindent
its retarded Green's function is given by

\begin{equation} 
G_R(x,x^{\prime}) = - \,  \theta (\eta - \eta^{\prime}) \,  G(x,x^{\prime})    
\label{Gret} 
\end{equation}

\noindent
Here $ \theta (\eta)  $ is the Heaviside step function, and $G(x,x^{\prime})$ 
is the Schwinger function for the field $\phi (\eta, \vec{x})$

\begin{equation} 
G(x,x^{\prime}) = - i \,  \langle 0 | \left[ \, \phi (x) \, , \, 
\phi (x^{\prime}) \right] | 0 \rangle     
\label{Schwingercomm} 
\end{equation}

Then, by inserting the mode expansion (\ref{scalphi}) in (\ref{Schwingercomm}), 
a straightforward computation yields

\begin{equation} 
G(x,x^{\prime}) = - \, \frac{ H^2 }{8\pi^2 \,  |\vec{x} - \vec{x}^{\prime} | } 
\, \left( \eta \eta^{\prime} \, I_1 - i\, (\eta - \eta^{\prime} ) \, 
I_2 + I_3  \right)     
\label{Schwingerint} 
\end{equation}

\noindent
where

\begin{equation} 
I_n = \int_{-\infty}^{\infty} \frac{du}{u^{n-1}} \left( \exp (iuy_-) - 
\exp (iuy_+) \right) ~~~;~~~ n = 1, 2, 3     
\label{integrals} 
\end{equation}

\noindent
and 

\begin{equation} 
y_\pm = \eta - \eta^{\prime}  \pm |\vec{x} - \vec{x}^{\prime}|   
\label{ymasmenos} 
\end{equation}

The integral $I_1$  gives 

\begin{equation} 
I_1 = 2\pi \left( \delta (y_-) -  \delta (y_+) \right)   
\label{Iuno} 
\end{equation}
 
\noindent
The integral $ I_2$ is clearly convergent, and its evaluation using residues 
theorem yields

\begin{equation} 
I_2 = 2\pi i \left( \theta (y_-) -  \theta (y_+) \right)   
\label{Idos} 
\end{equation} 

Notice that the integral $I_3$ is apparently logarithmically divergent in the 
infrared  -the integration variable being
$u = |\vec{k}|$- since  the  integrand behaves for small $u$ as

\begin{equation} 
\frac{1}{u^2} \,  \left(\, \exp (iuy_-) - \exp (iuy_+)\, \right) = 
- \frac{i}{u}\, 2 \, |\vec{x} - \vec{x^{\prime}}| + \cdots
\label{ucero} 
\end{equation}

However, this divergence does not really exist because  $I_3$ 
can be rewritten as

\begin{equation} 
I_3 = I_3^{\prime} + \hat{I_3}
\label{Isubtres}
\end{equation}

\vspace{-0.2cm}
\noindent
where

\begin{equation} 
I_3^{\prime} =  \int_{-\infty}^{\infty} \frac{du}{u^2} \, \left(\, 
\exp (iuy_-) - \exp (iuy_+) + 2i \, |\vec{x} - \vec{x}^{\prime}| 
\sin u \, \right)  
\label{Iprimatres}
\end{equation}

\noindent
and

\vspace{-0.2cm}

\begin{equation} 
\hat{I_3} = - 2i \,  |\vec{x} - \vec{x}^{\prime}|\,  \int_{-\infty}^{\infty} 
\frac{du}{u^2} \, \sin u  
\label{Igorrotres}
\end{equation}

Then $I_3^{\prime}$ is convergent, and $\hat{I_3} $ vanish -in the sense of 
Cauchy principal value- because the integrand is an odd function. Therefore, 
$I_3 = I_3^{\prime}$, and its evaluation using residues theorem, yields

\begin{equation} 
 I_3 = \pi \,  \left( \, | y_+| -   |y_-| \,  \right)    
\label{Itres}
\end{equation}

Thus, the Schwinger function $G(x,x^{\prime})$ is perfectly finite in the 
infrared limit, and replacing expressions
(\ref{Iuno}), (\ref{Idos}), and  (\ref{Itres})  in eq.  (\ref{Schwingerint}), 
it takes the form

\begin{eqnarray} 
G(x,x^{\prime}) = - \,  \frac{ H^2 }{4\pi \,  |\vec{x} - \vec{x}^{\prime}| }
+ \Big[  \, \eta \eta^{\prime} \,  \left(\delta (y_-) -  \delta (y_+)   \right)  
& + & (\eta - \eta^{\prime}) \, \left( \theta (y_-) -  \theta (y_+) \right)  
\nonumber \\
& + &  \frac{1}{2} \, \left(\, | y_+| -   |y_-| \,  \right) \, \Big]
\label{finSchwinger}
\end{eqnarray}

Finally, from eqs. ({\ref{Gret}) and ({\ref{finSchwinger}) we obtain the 
following simple expression for the de Sitter scalar
retarded Green's function 

\begin{equation} 
G_R(x,x^{\prime}) = \frac{H^2 \, \eta \eta{\prime}}{4\pi \,  
|\vec{x} - \vec{x}^{\prime}|} \, \delta \left( \eta - \eta^{\prime} -  
|\vec{x} - \vec{x}^{\prime} |\, \right)
+ \frac{H^2}{4\pi} \, \theta \left( \eta - \eta^{\prime} -  
|\vec{x} - \vec{x}^{\prime} |\, \right) 
\label{Greenscal}
\end{equation}

Notice that to obtain this expression, some cancellations have been produced 
in multiplying eq. (\ref{finSchwinger}) by $\theta (\eta - \eta^{\prime})$. 
This is due to the following identities among distributions

\begin{eqnarray} 
\theta (\eta - \eta^{\prime})\, \delta (y_+) &  = & 0 \nonumber \\
\theta (\eta - \eta^{\prime})\, \delta (y_-) &  = & \delta (y_-) \nonumber \\
\theta (\eta - \eta^{\prime})\, \theta (y_+) &  = & 
\theta (\eta - \eta^{\prime}) \nonumber \\ 
\theta (\eta - \eta^{\prime})\, \theta (y_-) &  = & \theta (y_-) \nonumber \\
\theta (\eta - \eta^{\prime})\, \left(\, | y_+| -   |y_-| \,  \right) &  = &
2\, (\eta - \eta^{\prime}) \theta (\eta - \eta^{\prime}) - 2\,  y_- \theta (y_-) 
\label{distident}
\end{eqnarray}

In particular, the advanced variable $ y_+ $ has disappeared from the arguments 
of the distributions entering (\ref{Greenscal}). Thus, the retarded Green's 
function (\ref{Greenscal}) has support on the past light cone and its interior, 
as it should be according to  the causality of the classical theory. 
Notice also that the main difference with the Minkowski retarded Green's 
function is the term proportional to $\theta(y_-)$. This term tell us that 
-although the free gravitational waves propagate  at the speed of light- 
in the production of gravity waves from sources, there is information about 
these sources, which propagates at a lower speed.
	
Expressions (\ref{GreenMink}) and (\ref{Greenscal}) completely solve the problem 
of the retarded graviton propagator in the de Sitter space-time. 
With this propagator at hand,  we are going to discuss in the next section the 
production of gravity waves by a generic source.
Notice that from now on we shall supress the upper label $(1)$  from  the 
energy-momentum tensor $T_{\mu \nu}$ of the source, in order to alleviate 
the notation.


\section{Gravitational waves produced by sources in de  Sitter space-time}
We proceed to obtain the gravitational field produced -in the linear 
approximation- by a generic source with energy-momentum tensor $T_{\mu \nu}$.  
From eqs. (\ref{GreenMink}), (\ref{eqGreenMink}) and (\ref{inhometachits}) 
the time-space components for this field are

\begin{eqnarray} 
\chi_{0 i}\left(\eta,\vec{x}\right) &  = & \eta \, 16 \pi\,  {\mathcal{G}}
\int \frac{d^3 \vec{x}^{\prime} \, d\eta^{\prime}}{\eta^{\prime}} \, \, 
G_R^{(M)}(x,x^{\prime}) \, T_{0i}(x^{\prime}) \nonumber \\
& = & 4 \, {\mathcal{G}}\, \eta  \int 
\frac{d^3 \vec{x}^{\prime}}{|\vec{x} - \vec{x}^{\prime} |}
\; \frac{1}{\eta - |\vec{x} - \vec{x}^{\prime}|}
\;T_{0i} (\eta - |\vec{x} - \vec{x}^{\prime} |\, , \, \vec{x}^{\prime}) 
\label{solts}	
\end{eqnarray}	

\noindent
Similarly, for the tilde component we have 

\begin{equation} 
\tilde{\chi} \left(\eta,\vec{x}\right) = 4 \, {\mathcal{G}} \, \eta  \int 
\frac{d^3 \vec{x}^{\prime}}{|\vec{x} - \vec{x}^{\prime} |}
\, \, \frac{1}{\eta  -  |\vec{x} - \vec{x}^{\prime} |} \;    
\tilde{T} (\eta - |\vec{x} - \vec{x}^{\prime} | \, , \, \vec{x}^{\prime}) 
\label{soltilde}	
\end{equation}

On the other hand, for the space-space components, eqs. (\ref{inhomchiss}), 
(\ref{simpeqscal}), and (\ref{Greenscal}) give

\begin{eqnarray}
\chi_{ij} \left(\eta,\vec{x}\right) &  = &
16 \pi\,  {\mathcal{G}} \int 
\frac{d^3 \vec{x}^{\prime} \, d\eta^{\prime}}{H^2 \eta^{\prime 2 }} \, \,  
G_R(x,x^{\prime}) \, T_{ij} (x^{\prime}) \nonumber \\
& = & 4 \, {\mathcal{G}} \int 
\frac{d^3 \vec{x}^{\prime}}{|\vec{x} - \vec{x}^{\prime} |}\, \,  
\frac{\eta}{\eta^{\prime}} \; \delta \left( \eta - \eta^{\prime} -  
|\vec{x} - \vec{x}^{\prime} |\, \right)
T_{ij} (\eta^{\prime}, \vec{x}^{\prime}) \nonumber \\
&  & + \;  4 \, {\mathcal{G}} \int d^3 \vec{x}^{\prime} \, d\eta^{\prime} \, \,  
\frac{1}{\eta^{\prime 2 }} \; \theta \left( \eta - \eta^{\prime} -  
|\vec{x} - \vec{x}^{\prime} |\, \right) 
T_{ij} (\eta^{\prime}, \vec{x}^{\prime})
\label{solss}	
\end{eqnarray}

\noindent
Then, using the identity 

\[
\frac{\eta}{\eta^{\prime}} \; \frac{1}{|\vec{x} - \vec{x}^{\prime} |} \;
\delta \left( \eta - \eta^{\prime} -  |\vec{x} - \vec{x}^{\prime} |\, \right) 
+ \frac{1}{\eta^{\prime 2}} \;  \theta \left( \eta - \eta^{\prime} -  
|\vec{x} - \vec{x}^{\prime} |\, \right)   =
\]

\vspace{-0.2cm}

\begin{equation} 
= \frac{1}{|\vec{x} - \vec{x}^{\prime} |} \; \delta \left( \eta - \eta^{\prime} 
- |\vec{x} - \vec{x}^{\prime} |\, \right) - \partial_{\eta^{\prime}}  
\left( \frac{1}{\eta^{\prime}} \; \theta \left( \eta - \eta^{\prime} 
- |\vec{x} - \vec{x}^{\prime} |\, \right)  \right)  
\label{identder}
\end{equation} 

\noindent
and doing a partial integration, eq. (\ref{solss}) can be recast as

\begin{eqnarray} 
\chi_{ij} \left(\eta,\vec{x}\right) & = & 
4 \, {\mathcal{G}}\, \int 
\frac{d^3 \vec{x}^{\prime}}{|\vec{x} - \vec{x}^{\prime}|} \;
T_{ij} (\eta - |\vec{x} - \vec{x}^{\prime} |\, , \, 
\vec{x}^{\prime}) \nonumber \\
&  & +\, 4 \, {\mathcal{G}}\,\int d^3 \vec{x}^{\prime} \, 
\int_{-\infty}^{\eta -  |\vec{x} - \vec{x}^{\prime}|}
 \; \frac{d\eta^{\prime}}{\eta^{\prime}}\; \partial_{\eta^{\prime}} 
T_{ij} (\eta^{\prime} \,  \vec{x}^{\prime})
\label{cojosolss}
\end{eqnarray}

Eqs. (\ref{solts}), (\ref{soltilde}), and (\ref{cojosolss}) give the linearized  
gravitational field $\chi_{\mu \nu}$ produced by a source with energy-momentum 
tensor $T_{\mu \nu}$ in de Sitter space-time. With this formulae at hand, we 
undertake now a double task: to do the spectral decomposition of the 
gravitational field $\chi_{\mu \nu}$ in frequencies, relating it to the spectral
decomposition of the source $T_{\mu \nu}$, and to show that this field takes the
form of radiated free gravitational waves in de Sitter space-time, when we go 
to the ``wave zone'' far away from the sources. For the first, we need an 
appropriate decomposition of the energy momentum tensor in modes of frequency 
$\omega$. For the space-space components, it is suitable to decompose the 
energy-momentum tensor in the form  

\begin{eqnarray} 
T_{ij} (\eta ,\vec{x}) & = & \int_{0}^{\infty}  d \omega \left[\, 
\exp (-i\omega \eta)  \left(\eta - \frac{i}{\omega} \right)
{\mathcal{T}}_{ij}(\omega ,\vec{x}) + 
\exp (i\omega \eta)  \left(\eta + \frac{i}{\omega} \right)  
{\mathcal{T}}^{\ast}_{ij}(\omega ,\vec{x})\, \right] \nonumber \\
& = & \int_{-\infty}^{\infty} d \omega  \exp (-i\omega \eta)  
\left(\eta - \frac{i}{\omega} \right) {\mathcal{T}}_{ij}(\omega ,\vec{x}) 
\label{decompT}
\end{eqnarray}

\noindent
i.e. we decompose $T_{ij}$ according to the modes entering the components 
$\chi_{ij}$ of a free gravitational wave. Notice that we have defined 
$ {\mathcal{T}}_{\mu \nu} (-\omega ,\vec{x}) \equiv 
{\mathcal{T}}^{\ast}_{\mu \nu}(\omega ,\vec{x})$ as usual. Now, since the 
modes $\exp (-i\omega \eta) \left(\eta - \frac{i}{\omega} \right)$ satisfy

\begin{equation}   
i \, \partial_{\eta} \exp (-i\omega \eta) \left(\eta - \frac{i}{\omega} 
\right) = \omega \eta \exp (-i\omega \eta)
\label{identinver}  
\end{equation}

\noindent
the integral transform (\ref{decompT}) can be easily inverted in the form

\begin{equation}   
{\mathcal{T}}_{ij}(\omega ,\vec{x}) = \frac{1}{2 \pi \, \omega} 
\int_{-\infty}^{\infty} d \eta\; \exp (i\omega \eta)
\, \frac{i}{\eta} \, \partial_\eta \, T_{ij} (\eta ,\vec{x})
\label{Fourinverss}
\end{equation}
 
For the other components of $T_{\mu \nu}$, since the corresponding  modes of 
a free gravitational wave take the form $\eta  \exp(i\omega \eta) $, 
we do the spectral decompositions

\begin{equation}   
T_{0i} (\eta ,\vec{x}) = \eta  \int_{-\infty}^{\infty} d \omega \; 
\exp (-i\omega \eta)\; {\mathcal{T}}_{ij}(\omega ,\vec{x})
\label{decompts}
\end{equation}

\vspace{-0.2 cm}
\noindent
and
\vspace{-0.2 cm}

\begin{equation}   
\tilde{T} (\eta ,\vec{x}) = \eta \int_{-\infty}^{\infty} d\omega\;  
\exp (-i\omega \eta)\; \tilde{{\mathcal{T}}}(\omega ,\vec{x})
\label{decomptilde}
\end{equation}

\noindent
whose inverses are

\begin{equation}  
{\mathcal{T}}_{0i} (\omega ,\vec{x}) =  \frac{1}{2\pi} 
\int_{-\infty}^{\infty} \frac{d \eta}{\eta} \,   \exp (i\omega \eta) \;   
T_{0i} (\omega ,\vec{x}) 
\label{Fourinverts}
\end{equation}

\vspace{-0.2 cm}
\noindent
and
\vspace{-0.2 cm}

\begin{equation}   
\tilde{{\mathcal{T}}} (\omega ,\vec{x}) =  \frac{1}{2\pi} 
\int_{-\infty}^{\infty} \frac{d \eta}{\eta} \,   \exp (i\omega \eta) \; 
\tilde{T}(\omega ,\vec{x})
\label{Fourinvertilde}
\end{equation}

\noindent
Now, let us decompose the gravitational field $\chi_{\mu \nu}$ in modes of 
frequency $\omega$ 

\begin{equation}
\chi_{\mu \nu}(\eta ,\vec{x}) =   \int_{0}^{\infty}  d \omega \;  
\chi_{\mu \nu }^{(\omega)}(\eta ,\vec{x})
\label{decompchimunu}
\end{equation}

\noindent
Then, for the space-space components of $\chi_{ij}$,  replacing the spectral 
decomposition (\ref{decompT}) for $T_{ij}$ in the expression  (\ref{cojosolss}), 
we have 

\begin{eqnarray}     
\chi_{ij}^{(\omega)}(\eta ,\vec{x}) &  =  & 4 \, {\mathcal{G}}\, \int    
\frac{d^3 \vec{x}^{\prime}}{|\vec{x} - \vec{x}^{\prime}|}   
\left[ \, \exp (-i\omega \eta)\, \exp (i\omega |\vec{x} - \vec{x}^{\prime}|)
\left(\eta - |\vec{x} - \vec{x}^{\prime}|  - \frac{i}{\omega} \right) \; 
{\mathcal{T}}_{ij}(\omega ,\vec{x}^{\prime}) + 
\mathrm{c.} \; \mathrm{c.} \, \right] \nonumber \\
& & + 4 \, {\mathcal{G}} \int  d^3 \vec{x}^{\prime}  
\int_{-\infty}^{\eta -  |\vec{x} - \vec{x}^{\prime}|}           
\left[ \,  - i\, \omega \,  {\mathcal{T}}_{ij}(\omega ,\vec{x}^{\prime})
\int_{-\infty}^{\eta -  |\vec{x} - \vec{x}^{\prime}|} d \eta^{\prime} 
\, \exp (-i\omega \eta^{\prime}) + \mathrm{c.} \; \mathrm{c.} \, \right] 
\label{pasodecompchiss} 
\end{eqnarray}

\noindent
and doing the usual shift $\omega \longrightarrow \omega + i \epsilon$, to 
handle the lower limit of the integral over $\eta^{\prime}$ in the second line 
of (\ref{pasodecompchiss}), we find

\begin{equation}
\chi_{ij}^{(\omega)}(\eta ,\vec{x}) = 4 \, {\mathcal{G}}\,
\left(\eta - \frac{i}{\omega} \right)   \exp (-i\omega \eta) 
\int \frac{d^3 \vec{x}^{\prime}}{|\vec{x} - \vec{x}^{\prime}|}\, 
\exp (i\omega |\vec{x} - \vec{x}^{\prime}|) \; {\mathcal{T}}_{ij}
(\omega ,\vec{x}^{\prime}) \; + \; \mathrm{c.} \; \mathrm{c.}
\label{cojodecompchiss}
\end{equation}

Similarly, replacing the spectral decompositions (\ref{decompts}) and 
(\ref{decomptilde}) for  $T_{0i}$ and $\tilde{T}$ in the expressions 
(\ref{solts}) and (\ref{soltilde}) for $\chi_{0i}$ and $\tilde{\chi}$ we have

\begin{equation}
\chi_{0i}^{(\omega)}(\eta ,\vec{x}) = 4 \, {\mathcal{G}}\, \eta \, 
\exp (-i\omega \eta) 
\int \frac{d^3 \vec{x}^{\prime}} {|\vec{x} - \vec{x}^{\prime}|}\, 
\exp (i\omega |\vec{x} - \vec{x}^{\prime}|)
\; {\mathcal{T}}_{0i}(\omega ,\vec{x}^{\prime}) \; + \; \mathrm{c.} \; \mathrm{c.}
\label{cojodecompchits}
\end{equation}

\noindent
and

\begin{equation}
\tilde{\chi}(\eta ,\vec{x}) = 4 \, {\mathcal{G}}\, \eta \,  \exp (-i\omega \eta) 
\int \frac{d^3 \vec{x}^{\prime}} {|\vec{x} - \vec{x}^{\prime}|}\, 
\exp (i\omega |\vec{x} - \vec{x}^{\prime}|) \; \tilde{{\mathcal{T}}} 
(\omega ,\vec{x}^{\prime}) \;  + \; \mathrm{c.} \; \mathrm{c.}
\label{cojodecompchitilde}
\end{equation} 

The expressions (\ref{cojodecompchiss}-(\ref{cojodecompchitilde}) 
are one of the main results of this paper. Just as in Minkowski space-time, 
these formulae relate the $\omega$ frequency component 
of the gravitational field produced by the source $T_{\mu \nu}$, with the 
$\omega$ frequency component of $T_{\mu \nu}$ itself. On the  other hand, 
the $\eta$-time dependent factors in front of the integrals in these formulae, 
exactly coincide with the $\eta$-time dependent factors for a free gravitational 
wave of frequency $\omega$. This is due to the appropriate decomposition of the 
energy momentum tensor which has been done in eqs. (\ref{decompT}), 
(\ref{decompts}), and (\ref{decomptilde}). This decomposition is of course
different of the  plain Fourier transformation which is done in Minkowski
space-time, since the form of the spectral transform is dictated by the form 
of the free gavitational waves themselves. Moreover,  we can  choose a source 
localized in a finite spatial region, and consider the gravitational field 
$\chi_{\mu \nu}$ produced by this source in the ``wave zone''. To do it, we go 
far away from the source and take the region of points $\vec{x}$ that satisfy

\begin{equation}
|\vec{x}| \gg \max \{ a \, ,\, \omega^2 a \}
\label{wavezone}
\end{equation}

\noindent
where $a$ is the size of the source. In this limit we have as usual

\begin{equation}
\frac{1}{|\vec{x} - \vec{x}^{\prime}|} \sim \frac{1}{r} ~~~,~~~ 
\exp (i\omega |\vec{x} - \vec{x}^{\prime}| ) \sim  
\exp (i \vec{k} \vec{x} - i \vec{k} \vec{x}^{\prime}) 
\label{wavezoneexp}
\end{equation}

\noindent
with $r = |\vec{x}|$, and $\vec{k} = \omega \, \vec{x}/ |\vec{x}|$. Thus, 
the expressions (\ref{cojodecompchiss}), (\ref{cojodecompchits}), and 
(\ref{cojodecompchitilde}) give in this limit

\begin{equation}
\chi_{ij}^{(\omega)}(\eta ,\vec{x}) \sim  \frac{ 4 \, {\mathcal{G}}}{r}\,
\left(\eta - \frac{i}{\omega} \right) \exp (-i\omega \eta + i \vec{k} \vec{x}) 
\, A_{ij} (\vec{k}) 
\label{ondass}
\end{equation}

\begin{equation}
\chi_{0i}^{(\omega)}(\eta ,\vec{x}) \sim  \frac{ 4 \, {\mathcal{G}}}{r} \, 
\eta \, \exp (-i\omega \eta + i \vec{k} \vec{x}) \,  A_{0i} (\vec{k}) 
\label{ondats}
\end{equation}

\begin{equation}
\tilde{\chi}^{(\omega)}(\eta ,\vec{x}) \sim  \frac{ 4 \, {\mathcal{G}}}{r} \, 
\eta \, \exp (-i\omega \eta + i \vec{k} \vec{x}) \,   \tilde{A} (\vec{k})
\label{ondatilde}
\end{equation}

\noindent
which correspond to free gravitational spherical waves in de Sitter space-time, 
being radiated away from the source, and whose amplitudes are given in terms of 
the energy-momentum tensor of the source by

\begin{equation}
A_{\mu \nu}(\vec{k}) = \int d^3 \vec{x}^{\prime} \, 
\exp (- i \vec{k} \vec{x}^{\prime}) \; 
{\mathcal{T}}_{\mu \nu} (\omega ,\vec{x}^{\prime})
\label{amplitudes}
\end{equation}

As a final remark, it is interesting to write down the spectral decomposition 
in frequencies of the covariant conservation law for the source
 energy momentum tensor 
$T_{\mu \nu}$. Using the expression (\ref{covdercov}) for the de 
Sitter covariant derivative, the conservation equation $D^\nu T_{\mu \nu} = 0$ 
splits into the two equations

\begin{equation}
- \partial_\eta T_{00} + \partial_j T_{0j} + \frac{1}{\eta} \, \tilde{T} = 0
\label{covconst}
\end{equation}

\vspace{-0.4cm}

\begin{equation}
- \partial_\eta T_{0i} + \partial_j T_{ji} + \frac{2}{\eta} \,  T_{0i}  = 0
\label{covconss}
\end{equation}

\noindent
Notice that these equations formally coincide with the gauge fixing conditions 
for $\chi_{\mu \nu}$ (\ref{chifixt}) and (\ref{chifixs}). Nevertheless, the 
geometrical meaning is not the same for both, because  $T_{\mu \nu}$ is a 
tensor while $\chi_{\mu \nu}$ is a tensor density. Now, using the spectral 
decompositions in frequencies (\ref{decompT}), (\ref{decompts}) and 
(\ref{decomptilde}), the conservation laws for the $\omega$ frequency
component of the energy-momentum tensor takes the form

\begin{equation}
i\, \omega \left( \tilde{\mathcal{T}} (\omega , \vec{x}) - 
{\mathcal{T}}_{ii} (\omega , \vec{x}) \right)
+ \partial_j {\mathcal{T}}_{0j} (\omega , \vec{x}) = 0 
\label{Fourconst}
\end{equation}

\vspace{-0.4cm}

\begin{equation}
i\, \omega \; {\mathcal{T}}_{0i} (\omega , \vec{x}) + \partial_j 
{\mathcal{T}}_{ji} (\omega , \vec{x}) = 0 
\label{Fourconss}
\end{equation}

\noindent
Notice that eq. (\ref{Fourconss}) has exactly the same form as in Minkowski 
space-time, while eq. (\ref{Fourconst}) is different because $\tilde{\mathcal{T}} 
(\omega , \vec{x}) - {\mathcal{T}}_{ii} (\omega , \vec{x}) \neq
{\mathcal{T}}_{00} (\omega , \vec{x})$, (in fact there is no suitable way 
to define $ {\mathcal{T}}_{00} (\omega , \vec{x}))$. Finally, using the 
conservation laws (\ref{Fourconst}) and (\ref{Fourconss}), one can check that 
the $\omega$ frequency component $\chi^{ (\omega) }_{\mu \nu}$ of the 
gravitational field $\chi_{\mu \nu}$, given by (\ref{cojodecompchiss}),
(\ref{cojodecompchits}) and  (\ref{cojodecompchitilde}), fulfil the gauge 
fixing conditions (\ref{chifixt}) and (\ref{chifixs}).


\section{Conclusions}
We have shown that the production of gravitational radiation from sources
moving in the 4-D de Sitter background can be studied along the same lines as
for Minkowski space-time. The  maximal symmetry and the conformal
flatness of the de Sitter space-time  are found  to be 
 two  key ingredients in order to achieve 
this goal. In addition, we have shown that -although the general equations 
for linear gravitational perturbations are rather cumbersome-  choosing the
gauge (\ref{psifix}),  the equations for all (physical and unphysical)  
polarizations of the graviton decouple, and amount to the equations for a 
 de Sitter minimally coupled  
massless scalar field  and a Minkowski massless field. 
In this respect, it is worth remarking that the minimally coupled 
massless  scalar field behaviour can be easily obtained, 
in the case of the physical
polarizations, by  imposing the traditional
synchronous transverse traceless gauge conditions
$D_\nu \psi_\mu^\nu = u_\nu \psi_\mu^\nu = 0$. However, these
two conditions cannot be simultaneously imposed in the presence of a source, 
and if one imposes
only the condition $D_\nu \psi_\mu^\nu = 0$ instead of (\ref{psifix}), 
one is led to a much more difficult
coupled graviton wave equation than (\ref{cojoeq}), whose
solutions  contain spurious complications. The same thing happens for the 
residual gauge invariance allowed by the gauge condition
 $D_\nu \psi_\mu^\nu = 0$.
While the infinitesimal coordinate transformations preserving the gauge 
condition (\ref{psifix}) are given by vector fields whose equations for  
the time and space
components decouple, and give very simple mode solutions; the vector fields 
corresponding to the residual gauge invariance allowed by 
 $D_\nu \psi_\mu^\nu = 0$, satisfy a much more difficult coupled system of 
partial differential equations.

The main new results of this paper are given in section VI. In that section 
we have shown that decomposing  the energy-momentum tensor of a given 
generic source $T_{\mu}^{\nu} (\eta, \vec{x})$ in frequencies, by  using  
a spectral transform dictated by the modes of the free gravitational waves
in the curved background,
we have very simple closed formulae relating  the $\omega$ frequency component
of the linearized gravitational field produded by the source with the
 transform $ {\mathcal{T}}_{\mu \nu} (\omega ,\vec{x})$ of the
energy-momentum tensor. We also
show that for localized sources, the produced gravitational field takes
the form of free gravitational waves in de Sitter space-time being radiated
away from the source. Thus, the generation of gravitational radiation 
by sources in de Sitter space-time ressembles very closely to the same process
in Minkowski space-time, the main difference being in the form of 
the energy-momentum frequency transform, which enter the formulae for the
amplitudes of the radiated waves.
       
         As a previous step we have also shown in section V, how the graviton 
retarded Green's function in the Sitter space-time 
-needed to solve the graviton wave equation- can be easily obtained 
using QFT techniques. The most prominent
feature of this retarded Green's function is that in addition to a 
delta function term in  the retarded time, it also contains a term 
proportional to the Heaviside step function of the retarded time.
This second term shows that the information about the sources in de 
Sitter space-time propagates not only at the speed of light but also
at a lower speed.

In  our opinion, it would be very interesting to apply
the general formulae (\ref{cojodecompchiss})-(\ref{cojodecompchitilde})
that we have derived in this paper,
 to sources that  could exist  during the inflationary
period of the universe. As a first example we have  work in progress
\cite{radstrings}
concerning string sources, whose equations of motion in de  Sitter space-time
have been solved in the case of a ring ansatz \cite{strings}.
   

\section{Acknowledgements:}
One of the authors (JR) would like to acknowledge the warm hospitality at 
DEMIRM (Observatoire de Paris), where this work was carried to completion  
during fall 1998, and the financial support of DIRECCION GENERAL
DE ENSE\~{N}ANZA SUPERIOR E INVESTIGACION CIENTIFICA of spanish MEC, 
which made this research stay possible.


\newpage

\begin{appendix}

\section{Appendix}
In this appendix we collect a number of formulae  for covariant derivatives, 
curvature tensors and d'Alembertians for de Sitter space-time,  which are used 
in this paper. The four dimensional de Sitter metric in conformal coordinates 
$x^{\mu} = (\eta, \vec{x})$ reads

\begin{equation}
ds^2 = \gamma_{\mu \nu} \,  dx^{\mu}dx^{\nu} = \frac{1}{H^2 \eta^2} \,  
(- d\eta^2 + d\vec{x}^2)
\label{dSmet}
\end{equation}

\noindent
 i.e.

\vspace{-0.2cm}

\begin{equation}
\gamma_{\mu \nu} = \frac{1}{H^2 \eta^2} \, \eta_{\mu \nu}  
\label{dSmeteta}
\end{equation}

\noindent
where  $\eta_{\mu \nu} = {\rm diag}(-,+,+,+) $ is the Minkowski metric, and 
$H$ is the Hubble constant.

Then the metric connection, can be written in these coordinates as

\begin{equation}
\Gamma^{(0)\nu}_{\mu  \lambda} = - \frac{1}{\eta} \left(\delta_\mu^0 \, 
\delta_\lambda^\nu + \delta_\lambda^0 \, \delta_\mu^\nu
+ \delta_0^\nu \,  \eta_{\mu \lambda} \right)   
\label{Criscero}
\end{equation}

\noindent
Thus, the covariant derivatives for covariant and contravariant vectors read

\begin{equation}
D_\mu V_\kappa = \partial_\mu V_\kappa + \frac{1}{\eta} \left( \delta_\mu^0 \, 
V_\kappa +  \delta_\kappa^0 \, V_\mu + 
\eta_{\mu \kappa} \, V_0  \right)   
\label{covdercov}
\end{equation}

\begin{equation}
D_\mu V^\nu = \partial_\mu V^\nu -  \frac{1}{\eta} \left( \delta_\mu^\nu \,  V^0 
+ \delta_\mu^0 \,  V^\nu + 
\delta_0^\nu \,  \eta_{\mu \lambda} \, V^\lambda  \right)   
\label{covdercontra}
\end{equation}

\noindent
In particular we have

\begin{equation}
D_\lambda V^\lambda = \partial_\lambda V^\lambda  -  \frac{4}{\eta} V^0    
\label{divercov}
\end{equation}

\noindent
and

\begin{equation}
D_\lambda   \psi_\mu^\lambda  = \partial_\lambda  \psi_\mu^\lambda  -  
\frac{4}{\eta} \,  \psi_\mu^0 +
\frac{1}{\eta} \,  \delta_\mu^0 \, \psi
\label{divercovpsi}
\end{equation}

Since the de Sitter space-time is maximally symmetric, the Riemann and 
Ricci tensors take the form

\begin{equation}
R^{(0)}_{\mu \kappa \lambda \rho} =  H^2  \left( \gamma_{\mu \rho} \, 
\gamma_{\kappa  \lambda} - \gamma_{\mu \lambda} \,  \gamma_{\kappa  \rho} \right) 
\label{Riemanncero}
\end{equation}
 
\noindent
and

\begin{equation}
R^{(0)}_{\mu \kappa } = - 3 H^2 \,   \gamma_{\mu \kappa}  
\label{Riccicero}
\end{equation}

In addition to these tensors,  we  need  the scalar, vector and tensor 
d'Alembertians, which can be computed using the expression (\ref{Criscero}) 
for the metric connection. For a scalar field $\phi$, we have the scalar 
d'Alembertian 

\begin{equation} 
\frac{1}{H^2 \eta^2} \,  D^{\lambda} D_{\lambda}\,  \phi = 
\left( \Box +\frac{2}{\eta} \,  \partial_{\eta} \right) \phi
\label{scaldal}
\end{equation}

\noindent
where $ \Box $ is the  Minkowski d'Alembertian

\begin{equation} 
\Box \equiv  - \partial_{\eta}^2 + \vec{\nabla}^2 
\label{Minkdal}
\end{equation}

\noindent
For a vector field $\xi_\mu$, the vector d'Alembertian is

\begin{equation} 
\frac{1}{H^2 \eta^2} \,  D^{\lambda} D_{\lambda}\, \xi_\mu  = 
\Box \,  \xi_\mu + \frac{2}{\eta} 
\left[\, \partial_\mu \xi_0 + \delta_\mu^0
\, \eta^{\alpha \beta } \, \partial_{\alpha} \xi_\beta \,  \right] +
\frac{1}{\eta^2} \left[\, 3 \, \xi_\mu  +  2 \, \delta_\mu^0 \, \xi_0 \, \right]
\label{vectordal}
\end{equation} 

\noindent
Finally, for a tensor field $\psi_\mu^\nu$, a long but straightforward 
computation yields

\begin{eqnarray}
\frac{1}{H^2\eta^2} \, D^{\lambda} D_{\lambda}\, \psi_\mu^\nu  & = & 
\Box\,  \psi_\mu^\nu  + \frac{2}{\eta} \left[\, \partial_\eta \psi_\mu^\nu 
+ \partial_\mu \psi_0^\nu - \eta^{\nu \kappa }\, \partial_\kappa  \psi_\mu^0  
\, \right] \nonumber \\
& & +  \frac{2}{\eta} \left[ \,  \delta_\mu^0 \, \eta^{\nu \alpha } \,  
\partial_\beta \psi_\alpha^\beta 
- \delta_0^\nu \, \partial_\beta \psi_\mu^\beta \, \right] \nonumber \\
& & + \frac{2}{\eta^2} \left[ \, \psi_\mu^\nu + 2 \, \delta_\mu^0 \, \psi_0^\nu 
+ 2 \, \delta_0^\nu \,  \psi_\mu^0  
- \delta_\mu^0 \, \delta_0^\nu \, \psi - \delta_\mu^\nu \, \psi_0^0 \,  \right]
\label{tensordal}
\end{eqnarray} 

\end{appendix}
 



\begin{thebibliography}{99}

\bibitem{Zeldovich}
Ya. B. Zeldovich. My Universe, (chapter 6 and refs. therein).
Harwood Acad. Publishers. Chur. 1992.

\bibitem{Grishuk}
L. P. Grishchuk.
Zh. Eksp. Teor. Fiz. {\bf 67}, 825, (1974).
Sov. Phys. JETP {\bf 40}, 409, (1975).

L. P. Grishchuk, Y.V. Sidorov 
Phys. Rev  {\bf D42}, 3413, (1990)

L. P. Grishchuk, M. Solokhin 
Phys. Rev  {\bf D43}, 2566, (1991)


\bibitem{FordParker}
L. H. Ford, L. Parker
Phys. Rev  {\bf D16}, 1601, (1977)

\bibitem{AllenJacob}
B. Allen, T. Jacobson
Comm. Math. Phys. {\bf 103}, 669, (1986)

\bibitem{AllenLN} 
B. Allen. {\em Gravitons in de Sitter space-time} in 
Field Theory, Quantum Gravity and Strings II.
Lecture Notes in Physics 280. 
Ed. by H.J. de Vega and N. Sanchez. Springer Verlag. Berlin: 1986

B. Allen, M. Turyn 
Nucl. Phys. {\bf B 292}, 813, 1987 


\bibitem{TsamisWoodPL}
N. C. Tsamis, R. P. Woodard.
Phys. Lett. {\bf B 292}, 269, (1992)

\bibitem{TsamisWoodComm}
N. C. Tsamis, R. P. Woodard 
Comm. Math. Phys. {\bf 162}, 162, (1994)

\bibitem{strings}
H.J. de Vega, A.V. Mikhailov, N. Sanchez
Teor. Mat. Fiz. {\bf 94}, 232, (1993)

H.J. de Vega, A.L.  Larsen, N. Sanchez
Nucl. Phys.  {\bf B 427}, 643, (1994)

\bibitem{Weinberg}
S. Weinberg. Gravitation and Cosmology. (Wiley, New York, 1972)

\bibitem{Miessner} 
C. W. Misner, K. S. Thorne, J. A. Wheeler. Gravitation. 
(Freeman, San Francisco, 1972)

\bibitem{GradRyz}
I. S. Gradshteyn, and I. M. Ryzhik. Table of Integrals, Series and Products.
(Academic Press, London, 1980)

\bibitem{radstrings}
H.J. de Vega, J. Ramirez, N. Sanchez.
Gravitational radiation from strings evolving in de Sitter space-time.
({\em In preparation}) 

\end{thebibliography}
\end{document}